\documentclass[aps,reprint,10pt,showkeys, superscriptaddress,showpacs]{revtex4-2}
\usepackage{multirow}
\usepackage{enumitem}
\usepackage{amsfonts} 
\usepackage{amsmath}
\usepackage{amssymb}
\usepackage{graphicx}
\usepackage{subfigure}
\usepackage{color}
\usepackage{enumerate}
\usepackage[]{natbib}
\usepackage{sidecap}
\usepackage{soul}
\usepackage{cancel}
\usepackage[normalem]{ulem}
\usepackage{dcolumn}%
\usepackage{bm}%

\newcommand*\xbar[1]{%
  \hbox{%
    \vbox{%
      \hrule height 0.5pt 
      \kern0.5ex
      \hbox{%
        \kern-0.1em
        \ensuremath{#1}%
        \kern-0.1em
      }%
    }%
  }%
} 

\begin{document}
    \title[]{ Gravitational instability in partially ionized plasmas: A two-fluid approach}
\author{Amar P. Misra}
\homepage{Author to whom any correspondence should be addressed}
\email{apmisra@visva-bharati.ac.in}
\affiliation{Department of Mathematics, Siksha Bhavana, Visva-Bharati University, Santiniketan-731 235, India}
 \author{Vinod Krishan}
\email{vinod@iiap.res.in}
\affiliation{Indian Institute of Astrophysics, Bangalore 560 034, India}

\begin{abstract}
We propose a new two-fluid model for a partially ionized magnetoplasma under gravity, in which electrons and neutrals are treated as a single fluid, while singly charged positive ions are a separate fluid. We observe that the classical result of gravitational instability (also known as Rayleigh-Taylor instability) in fully ionized plasmas becomes significantly modified by the influence of ion-neutral collisions (with frequency $\nu_{\rm{in}}$) and transverse wave numbers ($k_x$ and $k_y$). The instability growth rate can be enhanced or decreased depending on the values of the ratios $\kappa\equiv k_x/k_y$ and $f\equiv\nu_{\rm{in}}/\Omega_{\rm{ci}}$, where $\Omega_{\rm{ci}}$  is the ion-cyclotron frequency. We also estimate the growth rates relevant to the ionospheric E-region and solar atmosphere, noting that such growth rates can have a maximum for $\kappa,~f\ll1$, or for $\kappa>1$ and $f\sim0.64$, and minimized for $f\gg1$ irrespective of the value of $\kappa$. Furthermore, the timescale of instability ranges from $1$ minute to $2$ minutes in the solar atmosphere, while in the E region, it ranges from $1$ minute to $80$ minutes. The latter can be a satisfactory result for the reported lifetime of solar prominence threads.    
\end{abstract}
\maketitle
\section{Introduction} \label{sec-intro}
Partially ionized plasmas (PIPs) are ubiquitous in various astrophysical environments, including the solar atmosphere, accretion disks, interstellar medium, molecular clouds, stars, and cometary tails. The level of atomic ionization in these environments can range from nearly no ionization to full ionization, depending on whether the regions are cold or hot. Typically, the lower layers of the solar atmosphere are predominantly neutral. The presence of these neutral atoms leads to the formation of PIPs, whose main constituents are free electrons, ions, and neutral atoms [See, for a review article, e.g., Ref. \citep{ballester2018}]. Such PIPs are described using either a single-fluid or a multi-fluid model, of which a specific version is the two-fluid theory. In the former, the plasma loses the individual identities of its component species, and the single-fluid model applies to plasma environments where the collisional length and timescales are much larger than the scales of plasma oscillations. However, when such scales approach (or become smaller than) the characteristic scales for plasma oscillations, the single-fluid model breaks down. In that situation, a multi-fluid model, or, in particular, a two-fluid model, is required for the description of PIPs.
\par
Typically, strong electromagnetic fields can cause electrons and ions to move together, and, because of charge quasineutrality, they can move as a single fluid unit. Such modeling is well known in fully ionized plasmas. However, the charged species of PIPs would experience fundamentally different forces than neutral atoms. One realistic way to model PIPs is to construct a two-fluid system with charged species forming a single fluid (e.g., electrons and ions), and neutrals as a separate fluid. The reason is that in many PIP scenarios, collisions between electrons and ions can be as high as, or even higher than, ion-neutral collisions, allowing their momenta to be shared rapidly. These two fluids (electron-ion and neutral fluids) couple through collisions and ionizations/recombinations (See, e.g., Refs. \citep{hillier2023,soler2013}). 
\par 
On the other hand, it has been shown by Vranjes \textit{et al.} \citep{vranjes2015} that, when ion-neutral collision strongly dominates over similar other collisions, such as those in the lower solar atmosphere and lower Earth ionosphere (E-region), for low-frequency processes (compared to the collision frequency), ions and neutrals should be treated as a single fluid with an effective charge, while electrons are another fluid in PIPs. They have also shown that low-frequency magnetic waves can propagate in PIPs even if the particles are unmagnetized, and that because of charge exchange (also known as charge transfer or electron capture), separation of particles into two different populations as charged and neutral species may not be possible. Analogously, electrons accelerated by the Lorentz force always collide with neutral atoms, and this force can act as a drag force, transferring momentum from electrons to the neutral fluid. Also, since electrons have significantly smaller masses and higher speeds than ions, they can transfer momentum more frequently to neutrals through strong collisions.
\par
 There are three ways to model PIPs as a two-fluid system: (i) Combining appropriately electrons and neutrals to form a single fluid, the electron-neutral (EN) fluid, and ions as a second fluid, (ii) Combining ions and electrons as a single fluid and treating neutrals as the second fluid \citep{soler2022,diaz2012,ballai2019}, and (iii) Treating ions and neutrals as a single fluid and electrons as the second fluid. In the present letter, we explore the first possibility.
 Combining electrons and neutrals gives a rather novel system than combining ions and neutrals. Such an electron-neutral combination gives rise to a negatively charged fluid with a mass larger, smaller, or comparable to that of the ion fluid, forming a more general class of pair plasmas. On the other hand, combining ions and neutrals and treating electrons as another fluid is not significantly different from an electron-ion plasma, since the masses of ions and neutrals can be close to each other.    
  If the frequency of collision between electrons and neutrals is very high compared to the wave frequency, the two species can get highly coupled to behave as a single fluid with fluid mass density, $\rho_=\rho_e+\rho_n$, together with an effective charge and mass, to be defined. Here, $\rho_e~(\rho_n)$ is the fluid mass density of electrons (neutrals). Concurrently, if the collision frequency between ions and neutrals is low compared to the wave frequency, $\omega$, the ions and neutrals decouple, and ions can behave as a single fluid. The EN fluid and the ion fluid interact through electron-ion and ion-neutral collisions, as well as ionization and recombination processes. In the regions of PIPs, since the population of multiply charged ions is small, they can be assumed to be singly charged. Also, the magnetic field naturally influences both electrons and ions, but neutrals are not magnetized. Such PIPs are relevant to the solar atmosphere \citep{ballester2018,hillier2025,parenti2024}.
\par 
To justify the physical relevance of the model, we note that in the solar photosphere and chromosphere, the ionization fraction (electron-to-neutral density ratio) ranges from $10^{-6}$ to $10^{-3}$. The temperature ranges from $10^3$ to $10^4$ K  \citep{song2017}. Also, the collisional cross sections ($\sigma_{\rm{en}}$ and $\sigma_{\rm{in}}$) range from $10^{-19}$ to $10^{-18}~{\rm m}^2$ \citep{braileanu2019,vranjes2013}, and $T_{\rm e}\gtrsim T_{\rm i}$, $m_{\rm i}\gg m_{\rm e}$, and ${n_{\rm n0}}\gg{n_{\rm i0}}\sim n_{\rm e0}$. Here, $T_j$ denotes the thermodynamic temperature, $m_j$ the mass, and $n_{\rm j0}$ the unperturbed number density of the $j$-th species particle (j=e, i, and n, respectively, stand for electrons, ions, and neutrals).  In these environments, the ion-neutral $(\nu_{\rm in})$  and electron-ion $(\nu_{\rm ei})$ collision frequencies, compared to the electron-neutral collision frequency $(\nu_{\rm en})$, scale as 
${\nu_{\rm en}}/{\nu_{\rm in}}\sim \left({\sigma_{\rm en}}/{\sigma_{\rm in}}\right)\sqrt{{T_{\rm e}}/{T_{\rm i}}}\sqrt{{m_{\rm i}}/{m_{\rm e}}}$ (for $m_{\rm i}\ll m_{\rm n}$) and $ {\nu_{\rm en}}/{\nu_{\rm ei}}\sim \left({\sigma_{\rm en}}/{\sigma_{\rm ei}}\right) \left({n_{\rm n0}}/{n_{\rm i0}}\right)$,
where $\sigma_{\rm ij}$ denotes the corresponding collisonal cross sections. Typically, for the same background gas, the ratios, $\sigma_{\rm en}/\sigma_{\rm in}$ and $\sigma_{\rm en}/\sigma_{\rm ei}$ scale as $10^{-1}$. Also, $m_{\rm i}/m_{\rm e}\gtrsim 10^3$ and $n_{\rm n0}/n_{\rm e0},~n_{\rm n0}/n_{\rm i0}\sim10^3-10^6$ as mentioned before. Thus, we have either $\nu_{\rm en}>\nu_{\rm in}$, or $\nu_{\rm en}\gg\nu_{\rm in}$ (depending on the ratio $m_{\rm i}/m_{\rm e}\sim10^3$ or larger)  and $\nu_{\rm en}\gg\nu_{\rm ei}$. 
Thus, the collisional frequencies satisfy the following relations:
\begin{equation}\label{eq-coll-reg}
\nu_{\rm{ei}}\ll\nu_{\rm{en}};~\nu_{\rm in}<\nu_{\rm en},~\rm{or}~\nu_{\rm in}\ll\nu_{\rm en}.
\end{equation}
Also, $\nu_{\rm{in}}$ can be smaller, comparable to, or larger than the ion-cyclotron frequency, $\Omega_{\rm{ci}}\equiv eB_0/cm_{\rm i}$ with $e$ denoting the elementary charge, $B_0$ the static magnetic field, and $c$ the speed of light in vacuum.   
\par
In what follows, the ratio $R$ between the electron-neutral and ion-neutral collision  momentum transfer rates per unit volume scales as 
\begin{equation} \label{eq-moment-tran}
R\equiv\frac{m_{\rm e}\nu_{\rm en}|v_{\rm e}-v_{\rm n}|}{m_{\rm i}\nu_{\rm in}|v_{\rm i}-v_{\rm n}|}\sim \frac{\sigma_{\rm en}}{\sigma_{\rm in}}\sqrt{\frac{T_{\rm e}}{T_{\rm i}}}, 
\end{equation} 
where we have assumed that the ratio between the electron-neutral drift and the ion-neutral drift velocities scales as $\sim\sqrt{m_{\rm i}/m_{\rm e}}$ .  Typically, $\sigma_{\rm in}\gtrsim\sigma_{\rm en}$ and $T_{\rm e}\gtrsim T_{\rm i}$ as mentioned before. Thus, considering $\sigma_{\rm in}\sim 10\sigma_{\rm en}$, Eq. \eqref{eq-moment-tran} shows that the ratio $R$ can be larger than unity, i.e., the momentum transfer between electrons and neutrals becomes higher than that between ions and neutrals, for $T_{\rm e}> 10^2T_{\rm i}$. In this case, the higher electron-neutral momentum-transfer rate will yield a shorter mean-free path and tighter spatial coupling between electrons and neutrals, i.e., electrons will couple to neutrals at smaller scales than ions do, forming a single EN fluid, while ions remain a separate species. Thus, the coupling between electrons and neutrals occurs through electron-neutral collisions and higher momentum transfer, which forces neutrals to participate in the fluid motion under the magnetic force acting on electrons. Next, from the ratio between the electron-neutral collision and electron-cyclotron frequencies, we have 
\begin{equation} \label{eq-coll-cylo}
\frac{\nu_{\rm en}}{\Omega_{\rm ce}}=\frac{\sigma_{\rm en}}{\sigma_{\rm in}}\sqrt{\frac{T_{\rm e}}{T_{\rm i}}\frac{m_{\rm e}}{m_{\rm i}}}\left(\frac{\nu_{\rm in}}{\Omega_{\rm ci}}\right).
\end{equation}
Since $\sigma_{\rm in}\gtrsim\sigma_{\rm en}$, $T_{\rm e}\gtrsim T_{\rm i}$, $m_{\rm i}\gg m_{\rm e}$, and $\nu_{\rm{in}}$ can assume values smaller, close to, or larger than $\Omega_{\rm{ci}}$, the ratio in Eq. \eqref{eq-coll-cylo} can also be smaller than, comparable to, or larger than unity. In particular, ${\nu_{\rm en}}>{\Omega_{\rm ce}}$ holds, or the Lorentz force becomes weaker than the electron-neutral frictional force in the motion of electrons, for ${\nu_{\rm in}}>50{\Omega_{\rm ci}}$, i.e., for $f\equiv {\nu_{\rm in}}/{\Omega_{\rm ci}}\gg1$. 
 
\par 
We note that since the coupling processes between any charged species (e.g., electrons) and neutrals occur at very high frequencies, they have been hardly detected (or not detected) in observations historically. Such a limitation makes it difficult to directly study the significance of PIPs in the solar atmosphere. However, the advent of new observatories, e.g., Daniel K. Inouye Solar Telescope (DKIST), NASA's multi-slit Explorer (MUSE), could make it possible to provide a sufficient amount of observational data that will shed light on the small-scale high-frequency dynamics in the solar atmosphere, giving new insights into the physics of partially ionized solar plasmas. Thus, observations and modeling of PIPs are both crucial for new frontiers of solar plasma physics \citep{hillier2025}. 
\par 
Several authors have studied linear waves and instabilities in PIPs in astrophysical plasmas, including those in molecular clouds, solar atmosphere, and solar prominence \citep{soler2022,ballai2019,borah2007,kumar1990,bhatia1995}. The gravitational instability, or, more precisely, the Rayleigh-Taylor instability (RTI), can occur in plasmas, since the magnetic field acts as a lighter fluid to support heavier fluids like plasmas. Here, the centrifugal force on plasma particles in curved magnetic fields acts as an equivalent gravitational force. Several works in the literature focus on RTI in fully ionized plasmas (See, for some recent works, Refs. \citep{rozina2023,garai2020,rajaei2023,dey2025}). However, relatively less attention has been paid to investigating RTI in PIPs [See, e.g., Refs. \citep{diaz2012,diaz2014,ruderman2018,lukin2024}]. In the latter, the authors have either used a single-fluid approach or a two-fluid model, treating electrons and ions as a single fluid and neutrals as another. In Ref. \citep{diaz2012}, the authors studied RTI in a single interface between two partially ionized plasmas of different densities and considered electron-ion as a single fluid and neutrals as separate fluids. However, D{\' i}az \textit{et al.} \citep{diaz2014} reported a similar study on RTI in PIPs but using a single-fluid approach. In contrast, Ruderman \textit{et al.} \citep{ruderman2018} studied RTI at the interface between fully and partially ionized plasmas with the effects of magnetic shear. On the other hand, authors also reported nonlinear development of the magnetized RTI in a prominence-corona transition region \citep{lukin2024}.    
\par 
In this letter, in contrast to Refs. \citep{diaz2012,diaz2014,ruderman2018,lukin2024}, we consider the magnetic field to act as a lighter fluid to support heavier fluids like PIPs, in which  we assume electrons and neutrals are coupled to form a single fluid  with its own effective charge and mass, and ions as a separate fluid. 
 Such a two-fluid approach with the frequency regimes \eqref{eq-coll-reg} has not been considered before in the context of RTI or gravitational instability in PIPs.   
The coupling between electrons and neutrals occurs through electron-neutral collisions and momentum transfer, which force neutrals to participate in the fluid motion under the Lorentz force acting on both electrons and neutrals. 
\par 
The RTI (or, gravitational instability) for fully ionized two-fluid electron-ion magnetoplasmas under gravity is known in the literature (See, e.g., \citep{chen1983}). In its simplest version, for sufficiently small wave numbers (i.e., $kL_0\ll1$), the growth rate becomes
\begin{equation}
\gamma=\left[-g (n_0^{\prime}/n_0) \right]^{1/2}, \label{eq-gam-chen}
\end{equation}   
where $L_0=|n_0/n_0^{\prime}|$ is the plasma density inhomogeneity length scale, ${\bm g}=(g,0,0)$ is the constant gravitational acceleration, $n_0$ is the equilibrium plasma number density, and $n_0^{\prime}\equiv dn_0/dx~(<0)$ is the plasma density gradient along the $x$-axis. The main purpose of this work is to use the same formalism as in \citep{chen1983}, but for partially ionized magnetoplasmas under gravity that are relevant to astrophysical environments, including the solar atmosphere and the ionospheric E-region. We observe that the two transverse wave numbers ($k_x$ and $k_y$) and the ion-neutral collisions significantly modify the growth rate of RTI. 
\section{Equilibrium and basic equations} \label{sec-basic}
  We consider an unbounded, partially ionized magnetoplasma consisting of electrons, singly charged positive ions, and neutral atoms. We model this plasma as a two-fluid partially ionized plasma in which we appropriately combine the electron and neutral fluids by summation and difference of their respective fluid dynamical equations in the same manner as we obtain a single magnetohydrodynamic (MHD) fluid model by combining the electron and ion fluids in a fully ionized plasma. In this way, we get a negatively charged fluid with its own mass density $\rho$ and the center-of-mass velocity $\mathbf{v}$ defined by Eq. \eqref{eq-v-rho}. We study the gravitational instability (as described in Ref. \citep{chen1983}) by treating electrons and neutrals together as a single heavier fluid and ions as a separate lighter fluid. 
At equilibrium, there are flows of electrons, ions, and neutrals; the uniform magnetic field is along the $z$-direction, i.e., $\mathbf{B}=B_0\hat{z}$, and there is a plasma density gradient, $dn_{j0}/dx$ for $j$-th species particles, along the negative $x$ direction. We also assume that electrons and ions are cold, i.e., $T_e=T_i=0$, for simplicity, and the constant gravitational field in the positive $x$ direction, i.e., ${\bm g}=(g,0,0)$. This restriction on the field means we are interested in wavelengths much larger than the scale height for gravitational stratification. 
 \par 
 Next, we present the basic fluid equations for electrons, ions, and neutrals, namely the continuity and momentum balance equations. In various symbols, we use use Gaussian units throughout the text. Furthermore, we assume the mass relation among the species as $m_{\rm e}\ll m_{\rm i}\ll m_{\rm n}$. 
 \par 
 The momentum equations for electrons and neutrals, respectively, are
 \begin{equation}\label{eq-mom-e}
 \begin{split}
\rho_e\left[\frac{\partial}{\partial t}+\left({\bm v}_e\cdot\nabla\right)\right]{\bm v}_e=&-en_e\left({\bm E}+\frac1c{\bm v}_e\times {\bm B}\right)+\rho_e {\bm g} \\
&-\rho_e\sum_{j=i,~n}{\nu_{ej}\left({\bm v}_e-{\bm v}_j\right)},
\end{split}
\end{equation}
\begin{equation}\label{eq-mom-n}
\rho_n\left[\frac{\partial}{\partial t}+\left({\bm v}_n\cdot\nabla\right)\right]{\bm v}_n=\rho_n {\bm g}-\rho_n\sum_{j=i,~e}{\nu_{nj}\left({\bm v}_n-{\bm v}_j\right)},
\end{equation}    
where $\rho_j=n_jm_j$ and ${\bm v}_j$ are, respectively, the mass density (in which $n_j$ is the number density and $m_j$ the mass) and velocity of $j$-th species fluid, ${\bm E}$ is the electric field, and ${\bm B}$ is the uniform magnetic field. To present an equivalent momentum equation for the EN fluid, we combine Eqs. \eqref{eq-mom-e} and \eqref{eq-mom-n} and define the density and velocity for the EN fluid as follows.
\begin{equation} \label{eq-v-rho}
{\bm v}=\frac{1}{\rho}\left(\rho_e {\bm v}_e+\rho_n {\bm v}_n\right), ~\rho=\rho_e+\rho_n,~ {\bm u}={\bm v}_n-{\bm v}_e.
\end{equation}
In the limit of $\alpha_{\rm{e}}\equiv\rho_{\rm{e}}/\rho_{\rm{n}}\ll1$, we have ${\bm v}={\bm v}_n+\alpha_{\rm{e}} {\bm v}_e$, ${\bm v}_e\approx {\bm v}-{\bm u}$, and ${\bm v}_n\approx{\bm v}+\alpha_{\rm{e}} {\bm u}$.
\par 
 Next, combining Eqs. \eqref{eq-mom-e} and \eqref{eq-mom-n}, and noting that ${\bm v}\approx {\bm v}_n$ and $\rho\approx\rho_n $ for $\alpha_{\rm{e}}\ll1$, we obtain 
\begin{equation}\label{eq-mom-EN}
\begin{split}
\left(\frac{\partial}{\partial t}+ {\bm v}\cdot\nabla\right){\bm v}=&-\frac{e}{m_e}\alpha_{\rm{e}}\left[{\bm E}+\frac1c\left({\bm v}-{\bm u}\right) \times {\bm B}  \right] 
\\
&+{\bm g}-\nu_{ni}\left({\bm v}_n-{\bm v}_i\right)+\frac{{\bm v}_e}{\rho}\frac{\partial \rho_e}{\partial t},
\end{split}
\end{equation}
where we have used $\rho_e\nu_{\rm{en}}=\rho_n\nu_{\rm{ne}}$. When there is a significant background of neutrals, electron-neutral collisions become frequent. This creates a strong drag force that prevents the free Larmor gyration of electrons, in contrast to fully ionized plasmas. In this situation, the Lorentz force on the electron-neutral (EN) fluid outweighs that on electrons alone. Also, for the EN fluid to behave as a single fluid, we must have $u<v$. Therefore, we can neglect the term involving the relative velocity ${\bm u} \equiv {\bm v}_n - {\bm v}_e$ in Eq. \eqref{eq-mom-EN}. The term in this equation involving electron density variation (${\bm v}_e\partial\rho_e/\partial t$) is also less significant compared to the inertial term $\rho \partial {\bm v}/\partial t$ for EN fluids. Thus, neglecting the terms involving ${\bm u}$ and $\partial \rho_e/\partial t$ and considering $v_n\sim v$, Eq. \eqref{eq-mom-EN} can be represented as an ideal single EN-fluid equation as
\begin{equation}\label{eq-mom-en}
 \begin{split}
\left(\frac{\partial}{\partial t}+ {\bm v}\cdot\nabla \right){\bm v}=&-\frac{e}{m_e}\alpha_{\rm{e}}\left({\bm E}+\frac1c{\bm v}\times {\bm B}  \right) \\
&+{\bm g}-\nu_{ni}\left({\bm v}-{\bm v}_i\right).
\end{split}
\end{equation}
From Eq. \eqref{eq-mom-en}, it follows that the EN fluid experiences a reduced Lorentz force due to the appearance of the factor $\alpha_{\rm e}$. It is important to note that we have retained both the electron-neutral and ion-neutral drag forces in the equations of motion for electrons and neutrals, and the electron-neutral drag force cancels when adding these equations.  
\par 
The momentum balance equation for singly charged ions is
\begin{equation}\label{eq-mom-I}
 \begin{split}
&\left[\frac{\partial}{\partial t}+\left({\bm v}_i\cdot\nabla\right)\right]{\bm v}_i=\frac{e}{m_i}\left({\bm E}+\frac1c {\bm v}_i \times {\bm B}  \right)+{\bm g}\\
&-\nu_{\rm{in}}\left({\bm v}_i-{\bm v}_n\right)+\frac{\rho_e}{\rho_i}\nu_{\rm{ei}}\left({\bm v}_i-{\bm v}_e\right),
\end{split}
\end{equation}
where we have used $\rho_e\nu_{\rm{ei}}=\rho_i\nu_{\rm{ie}}$. We also note that $\nu_{\rm{ei}}\ll\nu_{\rm{en}}$, and the charge neutrality condition for electrons and ions holds for long-wavelength perturbations, i.e., $n_e\approx n_i$, so that $\rho_e\ll\rho_i$. Thus, neglecting the term proportional to $\rho_e/\rho_i$ in Eq. \eqref{eq-mom-I} and rewriting $v_n\sim v$, we have  
\begin{equation}\label{eq-mom-i}
 \begin{split}
\left[\frac{\partial}{\partial t}+\left({\bm v}_i\cdot\nabla\right)\right]{\bm v}_i=&\frac{e}{m_i}\left({\bm E}+\frac1c {\bm v}_i \times {\bm B}  \right)\\
&+{\bm g}-\nu_{\rm{in}}\left({\bm v}_i-{\bm v}\right).
\end{split}
\end{equation}
\par 
We also require the continuity equations for EN fluids and ions. These are given by
\begin{equation}\label{eq-con-en}
\frac{\partial \rho}{\partial t}+\nabla \cdot \left(\rho{\bm v}\right)=0,
\end{equation}
\begin{equation}\label{eq-con-i}
\frac{\partial \rho_i}{\partial t}+\nabla \cdot \left(\rho_i{\bm v}_i\right)=0.
\end{equation}
Equations \eqref{eq-mom-en}, \eqref{eq-mom-i}--\eqref{eq-con-i} are the desired set of equations for the EN and ion fluids in PIPs.
Before proceeding to perform the linearized wave analysis for the dispersion relation and the growth rate, we consider the equations for the equilibrium state as follows. From Eqs. \eqref{eq-mom-en} and \eqref{eq-mom-i}, we obtain  
 \begin{equation}\label{eq-mom-en0}
 -\frac{e\alpha_{\rm{e0}}}{cm_e}{\bm v}_0\times {\bm B}_0 +{\bm g}-\nu_{\rm{ni}}\left({\bm v}_0-{\bm v}_{i0}\right)=0,
\end{equation} 
\begin{equation}\label{eq-mom-i0}
\frac{e}{cm_i}{\bm v}_{i0} \times {\bm B}_0 +{\bm g}-\nu_{\rm{in}}\left({\bm v}_{i0}-{\bm v}_0\right)=0,
\end{equation}
where we have neglected the second-order terms. 
Taking the cross product of Eq. \eqref{eq-mom-en0} with ${\bm B}_0$ and separating the $x$ and $y$-components (Note that the equation corresponding to the $z$-component is identically satisfied) of the resulting equation, we obtain
\begin{equation} \label{eq-v0x}
-\alpha_{\rm{e0}}\Omega_{\rm{ce}}v_{\rm{0x}}+\nu_{\rm{ni}}\left(v_{\rm{0y}}-v_{\rm{i0y}}\right)=0,
\end{equation}  
 \begin{equation} \label{eq-v0y}
-\alpha_{\rm{e0}}\Omega_{\rm{ce}}v_{\rm{0y}}-\nu_{\rm{ni}}\left(v_{\rm{0x}}-v_{\rm{i0x}}\right)+g=0,
\end{equation}  
where $\alpha_{\rm{e0}}=m_{\rm e}n_{\rm{e0}}/m_{\rm n}n_{\rm{n0}}$.
Next, separating the $x$ and $y$-components of Eq. \eqref{eq-mom-i}, we obtain
\begin{equation} \label{eq-vi0x}
\Omega_{\rm{ci}}v_{\rm{i0y}}-\nu_{\rm{in}}\left(v_{\rm{i0x}}-v_{\rm{0x}}\right)+g=0,
\end{equation} 
 \begin{equation} \label{eq-vi0y}
\Omega_{\rm{ci}}v_{\rm{i0x}}+\nu_{\rm{in}}\left(v_{\rm{i0y}}-v_{\rm{0y}}\right)=0.
\end{equation}
In Eqs. \eqref{eq-vi0x} and \eqref{eq-vi0y}, the contribution of the term $\left({\bm v_{\rm i0}}\cdot \nabla\right) {\bm v_{\rm i0}}$ disappears, because we have considered the constant magnetic field ${\bm B}_0$ and the constant gravitational force, i.e., ${\bm g}$. So, to obtain the gravitational drift velocity for ions [from the ion momentum equation \eqref{eq-mom-i} at equilibrium], the velocity ${\bm v_{\rm i0}}$ and the collisional frequency $\nu_{\rm in}$ must be constant. In partially ionized inhomogeneous collision dominated plasmas, the particle mean-free path is much smaller than the spatial scale of density variation, i.e., over a mean-free path the density does not change significantly, resulting the ion-neutral collision frequency, $\nu_{\rm in}$, to be spatially constant and the term $\left({\bm v_{\rm i0}}\cdot \nabla\right) {\bm v_{\rm i0}}$ to vanish.   
Typically, in the case of gravitational instability or RTI, the gravitational force dominates over the frictional forces. Thus, from Eqs. \eqref{eq-vi0x} and \eqref{eq-vi0y}, we obtain
\begin{equation} \label{eq-vi0x1}
v_{\rm{i0x}}\approx\frac{g\nu_{\rm{in}} }{\Omega_{\rm{ci}}^2}\left(1+\frac{\nu^2_{\rm{in}}}{\Omega_{\rm{ci}}^2}\right)^{-1},
\end{equation} 
  \begin{equation} \label{eq-vi0y1}
v_{\rm{i0y}}\approx-\frac{g }{\Omega_{\rm{ci}}}\left(1+\frac{\nu^2_{\rm{in}}}{\Omega_{\rm{ci}}^2}\right)^{-1}.
\end{equation}  
In particular, in the absence of ion-neutral collisions, we have an agreement  with the classical result \citep{chen1983}: $v_{\rm{i0x}}=0$ and $v_{\rm{i0y}}=-g/\Omega_{\rm{ci}}$. Also, $v_{\rm{i0y}}=-\left(m_i/m_{\rm{en}}\right)g/\Omega_{\rm{cen}}$, i.e., $|v_{\rm{i0y}}|\ll g/\Omega_{\rm{cen}}$ for $m_i\ll m_{\rm{en}}$. Here, $\Omega_{\rm{cen}}\equiv |q_{\rm{en}}|B_0/cm_{\rm{en}}=\alpha_{\rm{e0}}\Omega_{\rm{ce}}$ is the EN-cyclotron frequency with $q_{\rm{en}}=-en_{\rm{e0}}/n_{\rm{n0}}$ denoting the effective charge and $m_{\rm{en}}\approx m_{\rm{n}}$ (since $\rho\approx \rho_{\rm{n}}$) the effective mass of EN fluid particles. By the same way as for the ion-drift velocities, we obtain from Eqs. \eqref{eq-v0x} and \eqref{eq-v0y} the following expressions for the EN-drift velocities.
\begin{equation} \label{eq-v0x1}
v_{\rm{0x}}\approx\frac{g\nu_{\rm{ni}} }{\Omega_{\rm{cen}}^2}\left(1+\frac{\nu^2_{\rm{in}}}{\Omega_{\rm{ci}}^2}\right)^{-1},
\end{equation} 
  \begin{equation} \label{eq-v0y1}
v_{\rm{0y}}\approx\frac{g }{\Omega_{\rm{cen}}}\left(1+\frac{\nu^2_{\rm{in}}}{\Omega_{\rm{ci}}^2}\right)^{-1},
\end{equation}  
where we have used $\rho_i\nu_{\rm{in}}=\rho_n\nu_{\rm{ni}}$ and the charge neutrality condition for electron and ion fluids, i.e., $n_{\rm{e0}}\approx n_{\rm{i0}}$.   
Equations \eqref{eq-vi0x1}-\eqref{eq-v0y1} are the new expressions for the drift velocities of ion and EN-fluid species in PIPs. We note that in the absence of ion-neutral collisions, the drift velocity of lighter species (ions) remains smaller than the drift velocity associated with heavier (EN) fluid, and it is reasonable to neglect the ion-drift velocities in the limit of $m_i/m_{\rm{en}}\to0$. Such an assumption agrees with the classical result of RTI \citep{chen1983}. Also, in contrast to fully ionized plasmas, the drift velocities of ion and EN species appear  both in the $x$- and $y$-directions. The $x$-component drifts appear due to ion-neutral collisions in the same direction, while ions and EN fluids have opposite drifts along the $y$-axis, similar to ion-electron plasmas \citep{chen1983}. In particular, in the absence of ion-neutral collisions, $v_{\rm{0x}}=0$ and $v_{\rm{0y}}=g/\Omega_{\rm{cen}}$. By comparing the relative magnitudes of the drift velocities, we find that for $m_i\ll m_n$ and $n_{\rm{e0}}\ll n_{\rm{n0}}$ (since in the solar photosphere and solar chormosphere, the ionization fraction varies from $10^{-6}$ to $10^{-3}$, \textit{cf}. Sec. \ref{sec-intro}), 
\begin{equation} \label{eq-coll-ratio}
\begin{split}
&\left|\frac{v_{\rm{i0x}}}{v_{\rm{0x}}}\right|,~\left|\frac{v_{\rm{i0y}}}{v_{\rm{0y}}}\right|\sim\frac{n_{\rm{e0}}}{n_{\rm{n0}}}\frac{m_i}{m_n}\ll1,\\
&\left|\frac{v_{\rm{i0x}}}{v_{\rm{i0y}}}\right|,~\left|\frac{v_{\rm{0x}}}{v_{\rm{0y}}}\right|\sim \frac{\nu_{\rm{in}}}{\Omega_{\rm{ci}}}. 
\end{split}
\end{equation}
From Eq. \eqref{eq-coll-ratio}, since $m_i\ll m_n$ and $n_{\rm{e0}}\ll n_{\rm{n0}}$, it follows that the drift velocities of the heavier EN fluid is always larger than that of the lighter ion fluid, and it is safe to neglect the contributions from $v_{\rm{i0x}}$ and $v_{\rm{i0y}}$ in the subsequent expressions for deriving the dispersion relation. Later, we will see that the drift velocity of the EN fluid contributes to the instability growth rate.  We also note that the $y$-component of the drift velocities can be larger or smaller than the $x$-components, depending on whether 
$\nu_{\rm{in}}$ is smaller or larger than $\Omega_{\rm{ci}}$. Typically, in the ionospheric E region, the ratio, $\nu_{\rm{in}}/\Omega_{\rm{ci}}$ varies from $0.72$ to $55$. In contrast, the same ratio varies between $0.01$  and $5$ in the solar chromosphere \citep{jiang2024}.  
\section{Dispersion relation and growth rate of instability} \label{sec-disp} %
To find the dispersion relation and the growth rate of instability, we perform the usual Fourier (linear) analysis for waves propagating in the $xy$ plane. 
 From Eq. \eqref{eq-mom-en}, the perturbed (quantities with suffix $1$) equation for EN fluid is
  \begin{equation}\label{eq-mom-en01}
 \begin{split}
m_{\rm n}&\left(n_{\rm n0}+n_{\rm n1}\right)\left[\frac{\partial}{\partial t}+ \left({\bm v}_0+{\bm v}_1\right)\cdot\nabla \right]\left({\bm v}_0+{\bm v}_1\right)= \\
&-e\left(n_{\rm e0}+n_{\rm e1}\right)\left[{\bm E}_1+
\frac1c\left({\bm v}_0+{\bm v}_1\right)\times {\bm B}_0\right]\\
&+m_{\rm n}\left(n_{\rm n0}+n_{\rm n1}\right)g
 -m_{\rm n}\nu_{\rm{ni}}\left({\bm v}_0+{\bm v}_1-{\bm v}_{\rm i0}-{\bm v}_{\rm i1}\right).
\end{split}
\end{equation}
Next, multiplying Eq. \eqref{eq-mom-en0} by $m_{\rm n}\left(n_{\rm n0}+n_{\rm n1}\right)$, subtracting the resulting equation  from Eq. \eqref{eq-mom-en01}, and neglecting the second-order terms, we obtain \citep{chen1983}  
 \begin{equation}\label{eq-mom-en1}
\left(\frac{\partial}{\partial t}+ {\bm v}_0\cdot\nabla \right){\bm v}_1=\frac{q_{\rm{en}}}{m_n}\left({\bm E}_1+
\frac1c{\bm v}_1\times {\bm B}_0\right) -\alpha_{\rm{i0}}\nu_{\rm{in}}{\bm v}_1,
\end{equation}
where $\alpha_{\rm{i0}}=\rho_{\rm{i0}}/\rho_{\rm{n0}}$, and we have neglected the contribution from the ion-velocity perturbation. This can be justified due to the fact that ion-drift velocities are much smaller than EN-drift velocities. 
\par 
 Next, assuming the perturbations to be of the form $\exp\left[i\left(k_xx+k_yy-\omega t\right)\right]$, where ${\bm k}=(k_x,k_y,0)$ is the wave vector and $\omega$ is the wave frequency, and separating the $x$- and $y$-components, we obtain in the limit of drift approximation \citep{chen1983}, $\Omega_{\rm{cen}}^2\gg \left(\omega-{\bm k}\cdot {\bm v}_0\right)^2$ (valid for strongly magnetized, low-frequency waves) the following expressions for the perturbed EN fluid velocities.
\begin{equation} \label{eq-v1xy}
v_{\rm{1x}}=\frac{c}{B_0}E_{\rm{1y}}, ~v_{\rm{1y}}=i\frac{\omega-{\bm k}\cdot {\bm v}_0}{\Omega_{\rm{cen}}}\frac{c}{B_0}E_{\rm{1y}}, 
\end{equation}
where we have neglected the electric field perturbation along the $x$-axis, i.e., $E_{\rm{1x}}=0$ \citep{chen1983}. This is justified since the ${\bm E}_1\times {\bm B}_0$-drift velocity is always upward or downward direction  (along the $x$-axis) depending on whether the surface of plasma waves move upward or downward.
In Eq. \eqref{eq-v1xy}, $v_{\rm{1y}}$ represents the polarization drift in the EN fluid frame, which is similar to the ion polarization drift (except the minus sign) in classical fully ionized plasmas \citep{chen1983}. 
\par 
By the same way, we obtain from Eq. \eqref{eq-mom-i} the following expressions for ions in the limit of $m_i/m_{\rm{en}}\to0$.
\begin{equation} \label{eq-vi1xy}
v_{\rm{i1x}}=\frac{c}{B_0}E_{\rm{1y}}=v_{\rm{1x}}, ~v_{\rm{i1y}}=0. 
\end{equation}
From Eq. \eqref{eq-con-en}, the perturbed equation of continuity for EN fluid is obtained as
\begin{equation}
\frac{\partial n_1}{\partial t}+n_0\nabla\cdot {\bm v}_1+{\bm v}_1\cdot\nabla n_0+{\bm v}_0\cdot\nabla n_1=0,
\end{equation}
which gives after Fourier wave analysis,
\begin{equation} \label{eq-n1}
\left(\omega-{\bm k}\cdot{\bm v}_0\right)n_1+\left(in_0^\prime-n_0k_x\right)v_{\rm{1x}}-n_0k_y v_{\rm{1y}}=0,
\end{equation}
where $n_0^\prime=dn_0/dx$ and $n\approx n_{\rm n}=n_{\rm n0}+n_{\rm n1}\equiv n_0+n_1$ for $\rho\equiv m_{\rm en}n\approx \rho_{\rm n}\equiv m_{\rm n}n_{\rm n}$ and $m_{\rm en}\approx m_{\rm n}$. 
\par 
Similarly, from the perturbed equation of continuity for ions, we obtain 
\begin{equation} \label{eq-ni1}
-i\omega n_{\rm{i1}}+\left(n^\prime_{\rm{i0}}+in_{\rm{i0}}k_x\right)v_{\rm{i1x}}=0,
\end{equation}    
 where $n_{\rm{i0}}^\prime=dn_{\rm{i0}}/dx$, and we have neglected the ion-drift velocities in view of Eq. \eqref{eq-coll-ratio}. 
 \par 
Substituting the expressions for $v_{\rm{1x}}$, $v_{\rm{1y}}$, $v_{\rm{i1x}}$, and $v_{\rm{i1y}}$ from Eqs. \eqref{eq-v1xy} and \eqref{eq-vi1xy} into Eqs. \eqref{eq-n1} and \eqref{eq-ni1}, and then assuming the plasma quasineutrality approximation, i.e.,  $n_1\approx n_{\rm{i1}}$ (valid for low-frequency, long-wavelength perturbations) and looking for nonzero solutions of the perturbed quantities, we obtain the following dispersion relation for low-frequency perturbations in PIPs.
  \begin{equation}\label{eq-disp1}
 \omega-{\bm k}\cdot{\bm v}_0-\left(B-\frac{k_yn_0}{\Omega_{\rm{cen}}}\frac{\widetilde{\omega}-{\bm k}\cdot{\bm v}_0}{n_0^\prime} \right)D\omega=0, 
\end{equation}
where $\widetilde{\omega}=\omega+i\alpha_{\rm{i0}} \nu_{\rm{in}}$, and the expressions for $B$ and $D$ are given by
\begin{equation} \label{eq-B-D}
B=1+\frac{n_0}{n_0^\prime}k_x,~ D=\left(\frac{n^\prime_{\rm{i0}}}{n_0^\prime}\right)^{-1} \left(1+i\frac{n_{\rm{i0}}}{n_{\rm{i0}}^\prime}k_x \right)^{-1}.
\end{equation}
From Eq. \eqref{eq-disp1}, the classical form of the dispersion relation \citep{chen1983} can be recovered for fully ionized electron-ion plasmas after proper substitutions and considerations, including $k_x=0$ and $n_{\rm{i0}}=n_0=n_{\rm{e0}}$.  Equation \eqref{eq-disp1} can be rewritten as
\begin{equation}
\omega^2-P\omega+Q=0,
\end{equation} 
where the coefficients $P$ and $Q$ are 
\begin{equation}\label{eq-P-Q}
\begin{split}
&P=\left(1-BD\right)\frac{\Omega_{\rm{cen}}}{k_yL_0D}+{\bm k}\cdot{\bm v}_0-i\alpha_{\rm{i0}}\nu_{\rm{in}},\\
&Q=\frac{\left({\bm k}\cdot{\bm v}_0\right)\Omega_{\rm{cen}}}{k_yL_0D},
\end{split}
\end{equation} 
and $L_0^{-1}=-n_0^\prime/n_0>0$ is the inverse of the length scale of EN-fluid density inhomogenity along the negative $x$-axis.                                 
\par 
Assuming $|\omega|\gg\alpha_{\rm{i0}} \nu_{\rm{in}}$ [\textit{cf}. Eq. \eqref{eq-coll-reg}], and looking for long-wavelength perturbations with $k^2_xL^2_0,~k^2_xL^2_{\rm{i0}}\ll1$, where $L_{\rm{i0}}^{-1}=-n_{\rm{i0}}^\prime/n_{\rm{i0}}>0$ is the inverse of the length scale of ion fluid density inhomogenity along the negative $x$-axis, so that $B\approx1$ and $D\approx \left(n_0^\prime/n^\prime_{\rm{i0}}\right)= \left(n_0/n_{\rm{i0}}\right)\left(L_{\rm{i0}}/L_0\right)\equiv D_0$, we obtain the real wave frequency and the instability growth rate as 
\begin{equation}\label{eq-re-freq}
\Re\omega\equiv \omega_{\rm r}=\frac12P_0\equiv\frac12\left[\left(1-D_0\right)\frac{\Omega_{\rm{cen}}}{k_yL_0D_0}+{\bm k}\cdot{\bm v}_0  \right],
\end{equation} 
\begin{equation} \label{eq-im-freq}
\Im\omega\equiv\gamma\approx Q_0^{1/2}=\left[\frac{\left({\bm k}\cdot{\bm v}_0\right)\Omega_{\rm{cen}}}{k_yL_0D_0}\right]^{1/2},
\end{equation}
provided 
\begin{equation} \label{eq-kyL0}
k_y^2L_0^2\ll \frac{4}{D_0}\frac{\Omega^2_{\rm{cen}}}{\omega_g^2}\left(1+\frac{\nu^2_{\rm{in}}}{\Omega^2_{\rm{ci}}}\right)\Big/\left(1+\frac{k_x}{k_y}\frac{\nu_{\rm{in}}}{\Omega_{\rm{ci}}}\right),
\end{equation}
where $\omega_g=\sqrt{g/L_0}$ is the Brunt-V{\"a}is{\"a}l{\"a} frequency. Typically, $D_0\lesssim1$ and $\Omega_{\rm{ce}}\gg\omega_g$ for solar atmospheric conditions, so that for $\alpha_{\rm{e0}}\ll1$, $\alpha_{\rm{e0}}\Omega_{\rm{ce}}\equiv \Omega_{\rm{cen}}\sim\omega_g$. Also, the ratio of the two factors involving $\nu_{\rm{in}}$ in the parentheses can be smaller than unity for $\kappa\equiv k_x/k_y>1$ and $0\lesssim f\equiv{\nu_{\rm{in}}}/{\Omega_{\rm{ci}}}\lesssim1$, or larger than unity for $\kappa<1$ and $0\lesssim f\lesssim50$. Thus, the term on the right-side of Eq. \eqref{eq-kyL0} can assume more or less a constant value, and the long-wavelength condition, $k_y^2L_0^2\ll1$ can be fulfilled. By means of Eqs. \eqref{eq-v0x} and \eqref{eq-v0y}, Eqs. \eqref{eq-re-freq} and \eqref{eq-im-freq} can be recast as 
\begin{equation} \label{eq-re-freq1}
\begin{split}
\omega_{\rm r}=&\frac{1}{2}\left[\frac{1-D_0}{D_0k_yL_0}  
 +k_yL_0\left(1+\frac{k_x}{k_y}\frac{m_{\rm i}}{m_{\rm en}} \frac{n_{\rm i0}}{n_0} \frac{\nu_{\rm in}}{\omega_{\rm g}} \right)\right. \\
 &\left.\times\left(1+\frac{\nu^2_{\rm{in}}}{\Omega^2_{\rm{ci}}}\right)^{-1} \right],
\end{split}
\end{equation}
\begin{equation} \label{eq-im-freq1}
\gamma=\left[\frac{g}{\widetilde{L}_0}\left(1+\frac{\nu^2_{\rm{in}}}{\Omega^2_{\rm{ci}}}\right)^{-1}\left(1+\frac{k_x}{k_y}\frac{\nu_{\rm{in}}}{\Omega_{\rm{ci}}}\right)\right]^{1/2},
\end{equation}
where $\widetilde{L}_0\equiv L_0D_0=-n_0/n^\prime_{\rm{i0}}$ is the effective length scale of plasma density inhomogeneity. Since $n_0\gg n_{\rm i0}$, this length scale appears to be higher than the classical result $\left(n_{\rm i0}/n^\prime_{\rm{i0}}\right)$  \citep{chen1983}, indicating a reduction of the growth rate [see Eq. \eqref{eq-im-freq1}] compared to the classical result with $\nu_{\rm in}=0$.  
From Eqs. \eqref{eq-re-freq} and \eqref{eq-im-freq1}, a similar form of the classical result can be recovered by setting $D_0=1$, $k_x=0$, and $\nu_{\rm{in}}=0$. Equation \eqref{eq-im-freq1} is the new result for Raleigh-Taylor instability in the two-fluid model of partially ionized magnetoplasmas. Since the ratio $f\equiv\nu_{\rm{in}}/\Omega_{\rm{ci}}$ can be of the order of unity or more in the environments of ionospheric E-region ($f\sim0.72-55$) or solar chromosphere ($f\sim0.88-2.2$) \citep{jiang2024}, the instability growth rate can be lower compared to the classical result by the influence of ion-neutral collisions. Interestingly, since the RTI occurs for ${\bm k}\perp {\bm B}_0$, the modification of the growth rate still occurs by the presence of $\nu_{\rm{in}}$  even if the perturbation along the $x$-axis is ignored (i.e., $k_x=0$). As mentioned before, for solar atmospheric conditions, $D_0\lesssim1$. Also, $n_0\gg n_{\rm i0}$, $m_i\ll m_{\rm en}$, and $\nu_{\rm in}$ can be smaller or larger than $\omega_{\rm g}$. So, Eq.          \eqref{eq-re-freq1} further reduces to
\begin{equation} \label{eq-re-freq2}
\omega_{\rm r}\approx\frac{1}{2}\left[\frac{1-D_0}{D_0k_yL_0} 
 +k_yL_0\left(1+\frac{\nu^2_{\rm{in}}}{\Omega^2_{\rm{ci}}}\right)^{-1} \right],
\end{equation}
In the absence of the ion-neutral collision, Eq. \eqref{eq-re-freq2} agrees with the classical result \citep{chen1983}. We numerically analyze the real wave frequency and the growth rate of instability for the parameters relevant to the ionospheric E region and solar atmosphere. 
\par
\begin{figure}[!h]
\centering
\includegraphics[width=9cm,height=5cm]{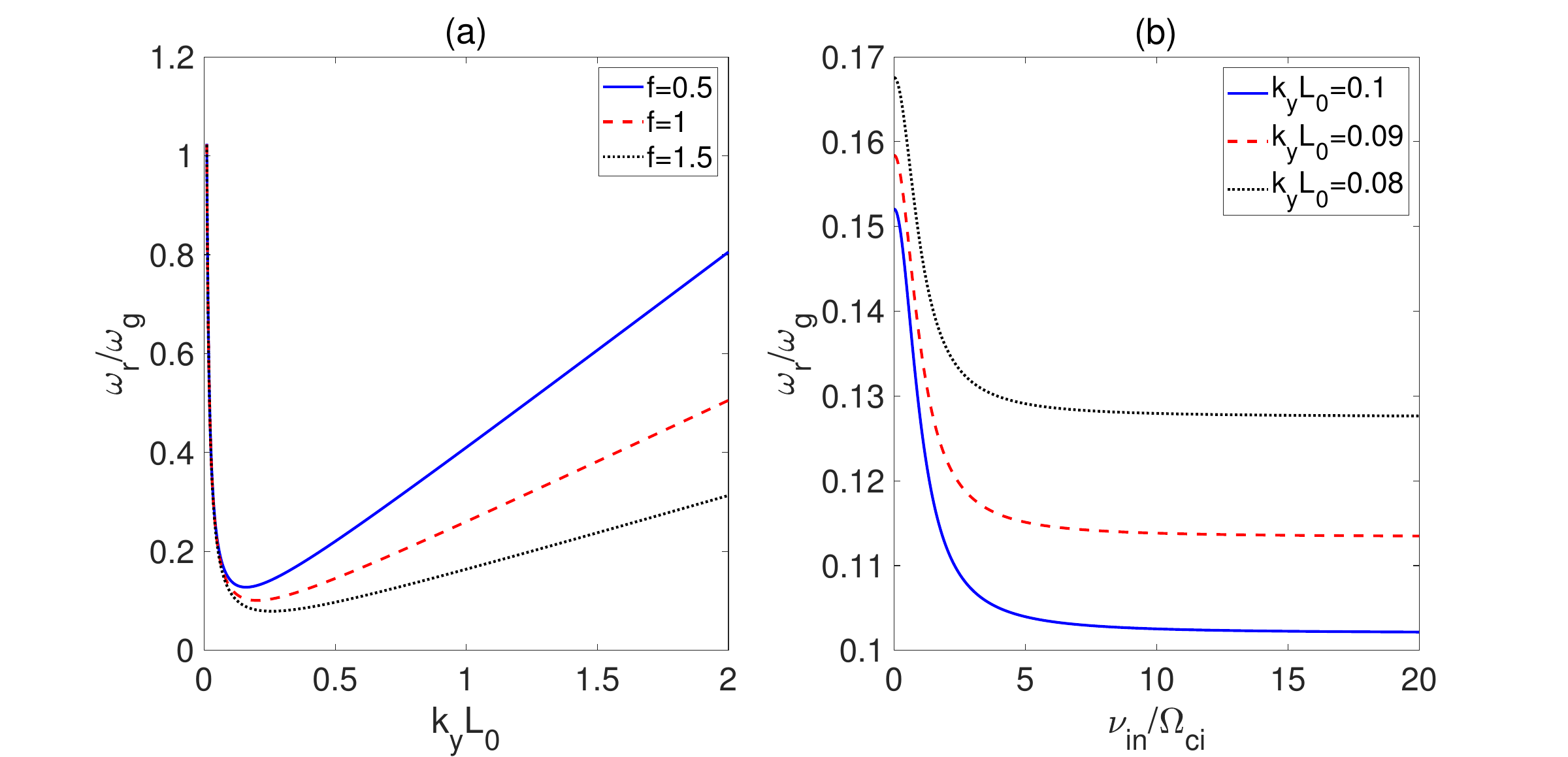}
\caption{The normalized real wave frequency [$\omega_{\rm r}/\omega_g$, see Eq. \eqref{eq-re-freq1}] is shown against the normalized wave number $k_yL_0$ [subplot (a)] and the frequency ratio $f\equiv\nu_{\rm{in}}/\Omega_{\rm{ci}}$ [subplot (b)] for different values of the parameters as in the legends. The fixed parameter values for the subplots (a) and (b) are $\kappa\equiv k_x/k_y\sim1$, $D_0\sim0.98$, and $n_{\rm i0}/n_{0}\sim 10^{-3}$. }
\label{fig1}
\end{figure}

 
Figure \ref{fig1} shows that for $D_0\sim1$, the real wave frequency initially drops, reaches a minimum value, and then increases with increasing values of the transverse wave number $k_yL_0$, indicating that while the real wave mode is dispersive for small $k_yL_0$, it tends to become dispersionless as $k_yL_0$ increases [subpplot (a)]. However, for a fixed value of $k_yL_0$, the wave frequency decreases and approaches a steady state value as the ratio, $\nu_{\rm{in}}/\Omega_{\rm{ci}}$ increases [subplot (b)].    
\par
\begin{figure}[!h]
\centering
\includegraphics[width=9cm,height=5cm]{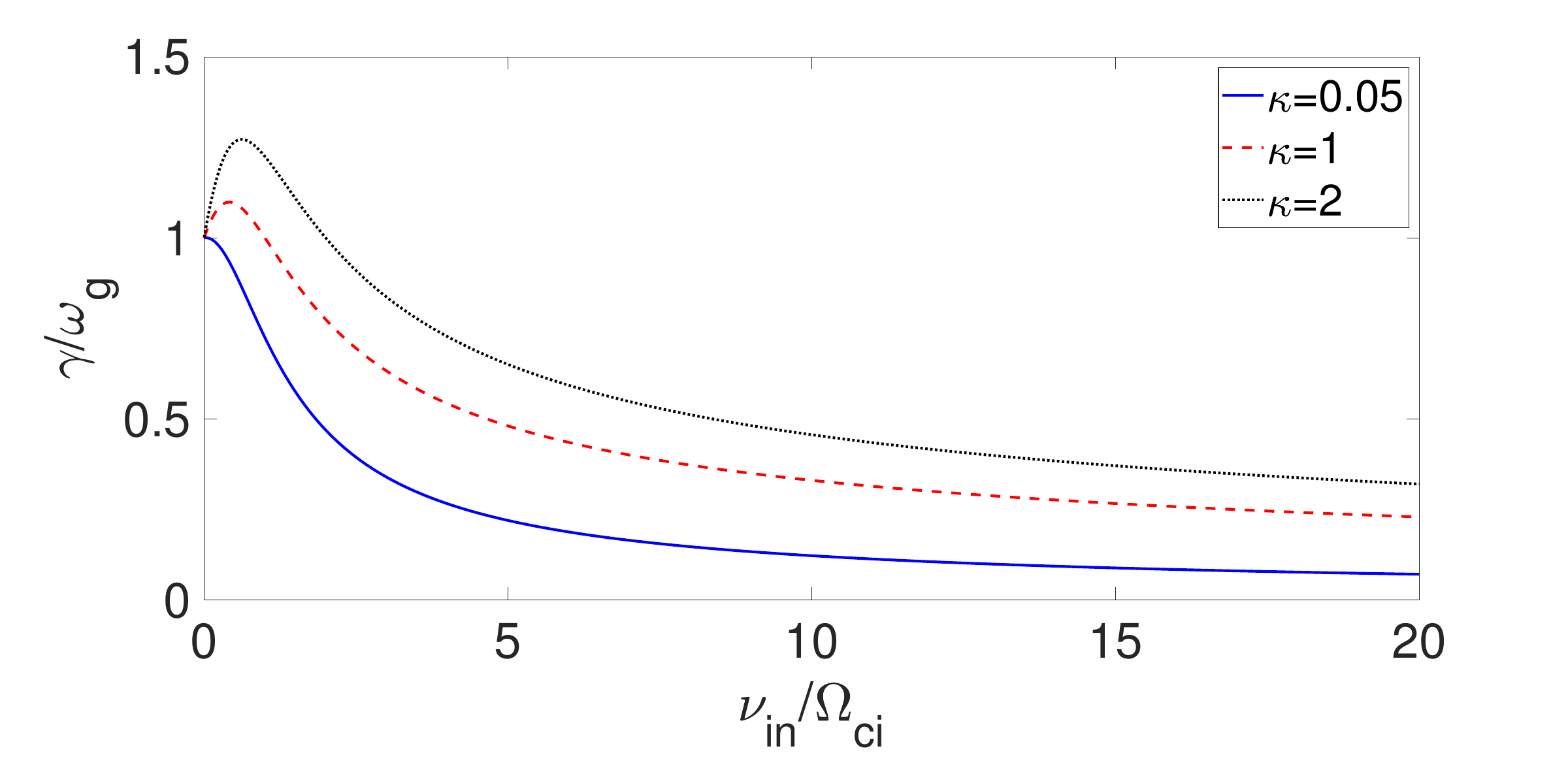}
\caption{The normalized growth rate of instability [$\gamma/\omega_g$, see Eq. \eqref{eq-im-freq1}] is shown against the frequency ratio $f\equiv\nu_{\rm{in}}/\Omega_{\rm{ci}}$ for different values of the wave-number ratio, $\kappa\equiv k_x/k_y$ as in the legends. }
\label{fig2}
\end{figure}

 From Eq. \eqref{eq-im-freq1}, it is evident that for a fixed value of $k_x$ ($k_y$), the growth rate always decreases (increases) as the wave number $k_y$ ($k_x$) increases. Also, depending on the ratio $\kappa\equiv k_x/k_y$, smaller or larger than unity, the growth rate can decrease or increase even when the frequency ratio  $f$ increases.
  From Fig. \ref{fig2}, we observe that for $\kappa<1$, i.e., when the wavelength of perturbation along the $y$-axis is small (or ignored) compared to the $x$-axis, the growth rate always decreases with increasing values of $f$. However, when $f\gtrsim1$, the growth rate increases initially in the interval $0<f\lesssim0.6$, but decreases for $f>0.6$. 
  \par 
    Typically, in the ionospheric E region \citep{kelley1989}, the ratio $f$ varies as $f\sim0.72-55$, while in the solar chromosphere ($800-1500$ km above the photosphere), we have \citep{vernazza1981} $f\sim0.01-5$. On the other hand, in laboratory experiments \citep{jiang2024}, the frequency ratio $f$ can vary, respectively, as $f\sim0.88-1.27$ and $1.40-2.20$ in weakly and strongly collisional plasmas. If the density-drop-scale size is $L_0\sim10^3$ km and the gravity acceleration is $g=274~\rm{m}/\rm{s}^{-2}$ (e.g., in the solar photosphere/chromosphere), we can obtain some estimates for the wave frequency and growth rate for different values of $\kappa$ or $k_{\rm y}L_0$. For example, for a fixed $k_{\rm y}L_0=0.1$, the frequency reaches a maximum value, $\omega^{\max}_{\rm r}\approx0.152\omega_{\rm g}\sim0.0025~\rm{s}^{-1}$ at $f\sim0$ and a minimum value $\omega^{\min}_{\rm r}\approx0.102\omega_{\rm g}\sim0.0017~\rm{s}^{-1}$ for $f\gg1$. Similarly,  for $\kappa\sim0.05$, the maximum growth rate is achieved at $f\sim0$, i.e., $\gamma_{\max}\sim\omega_g\sim0.0166~\rm{s}^{-1}$, and the minimum growth rate can be achieved for $f\gg1$ as $\gamma_{\min}\sim0.12\omega_g\sim2\times10^{-4}~\rm{s}^{-1}$. On the other hand, for $\kappa\sim2$, we have $\gamma_{\max}\sim1.27\omega_g\sim0.02~\rm{s}^{-1}$ at $f\sim0.64$ and $\gamma_{\min}\sim0.46\omega_g\sim0.0076~\rm{s}^{-1}$ for $f\gg1$. Thus,  one can conclude that in order to achieve higher growth rates, one must consider the perturbations both in $x$ and $y$ directions, i.e., transverse to the magnetic field, and the time scale of instability ranges from approximately $\tau_{\rm{RTI}}\sim1$ minute to $80$ minutes (relevant to the ionospheric E region) or $\tau_{\rm{RTI}}\sim1$ minute to $2$ minutes (relevant to the solar chromosphere). However, concerning the solar prominence threads,  while $\tau_{\rm{RTI}}\sim1$ is much shorter than the reported lifetime of the solar prominence threads \citep{lin2011}, $\tau_{\rm{RTI}}\sim80$ may be close to the lifetime of threads \citep{diaz2012}. Our results also indicate that collisions can significantly suppress the RTI, approaching a steady-state value. However, this is only a linear prediction; nonlinear results from a more advanced model may be necessary to verify it.  
\section{Conclusion} \label{sec-conclu}
 We have studied the gravitational instability, i.e., RTI, for low-frequency, long-wavelength electrostatic waves in partially ionized magnetoplasmas using a new two-fluid model in which electrons and neutrals behave as a single fluid with an effective charge and mass, and ions as a separate fluid. The two fluids interact via electron-neutral and ion-neutral collisions. In contrast to classical RTI in fully ionized plasmas \citep{chen1983}, a new length scale of plasma density inhomogeneity appears, leading to modification in the Brunt-V{\"a}is{\"a}l{\"a} frequency. We have shown that the growth rate of instability is significantly modified by ion-neutral collisions and wave propagation transverse to the magnetic field, thereby altering the results reported in the literature. The growth rate can be maximized either for $\kappa\equiv k_x/k_y,~f\equiv \nu_{\rm{in}}/\Omega_{\rm{ci}}\ll1$, or when $\kappa>1$ and $f\sim0.64$. However, it minimizes at $f\gg 1$ regardless of $\kappa$. We have estimated that the time scales of instability can range from approximately $\tau_{\rm{RTI}}\sim1$ minute to $80$ minutes, such as in the ionospheric E region, or from $\tau_{\rm{RTI}}\sim1$ minute to $2$ minutes, relevant to the solar chromosphere. Interestingly, $\tau_{\rm{RTI}}\sim80$ is close to the reported lifetime of the solar prominence threads \citep{diaz2012}. Thus, our results are also applicable to PIPs in solar prominences. However, we have considered a simple model and performed a linear analysis; therefore, nonlinear treatments with a more advanced model may be necessary to verify our prediction \citep{lukin2024}. Nevertheless, the present linear analysis in the EN-ion two-fluid model will provide guidelines for advancing existing theories in the literature and verifying future observational data in solar PIPs.     
\section*{Acknowledgments}
 V. Krishan gratefully acknowledges the valuable support received from the Department of Mathematics, Visva-Bharati, Santiniketan, during her visit. 
\subsection*{Declaration of Competing Interest}
The authors declare that they have no known competing financial interests or personal relationships that could have appeared to influence the work reported in this paper.
\subsection*{Author contributions}
 Amar Prasad Misra: Conceptualization (equal); Formal analysis (equal); Investigation (equal); Methodology (equal); Validation (equal); Writing--original draft.
Vinod Krishan: Conceptualization (equal); Formal analysis (equal); Investigation (equal); Methodology (equal); Validation (equal);  Writing--review \& editing.
\subsection*{Data availability} The data used in the manuscript are available at \citep{song2017,jiang2024,kelley1989}. 
 \section*{Conflict of interest}
The authors have no conflicts to disclose.
\bibliographystyle{apsrev4-2}
\bibliography{reference}

@book{chen1983,
 author = {Chen, F. F.},
 title = {TIntroduction to plasma physics and controlled fusion},
 publisher = {Plenum press},
 year = {1984},
 volume={1},
 pages={215},
 address = {New York}  
  }

@ARTICLE{soler2022,
 AUTHOR={Soler, Roberto  and Ballester, José Luis },
 TITLE={Theory of Fluid Instabilities in Partially Ionized Plasmas: An Overview},
 JOURNAL={Frontiers in Astronomy and Space Sciences},
 VOLUME={Volume 9 - 2022},
YEAR={2022},
DOI={10.3389/fspas.2022.789083},
ISSN={2296-987X}
}

@article{hillier2025,
    author = {Hillier, Andrew S. and Snow, Ben and Luna, Manuel},
    title = {Partially ionized plasma of the solar atmosphere: recent advances and future pathways},
    journal = {Philosophical Transactions of the Royal Society A: Mathematical, Physical and Engineering Sciences},
    volume = {382},
    number = {2272},
    pages = {20230230},
    year = {2024},
    month = {04},
    issn = {1364-503X},
    doi = {10.1098/rsta.2023.0230},
    url = {https://doi.org/10.1098/rsta.2023.0230}
    }

@article{ballester2018,
author={Ballester, Jos{\'e} Luis and Alexeev, Igor and Collados, Manuel and Downes, Turlough and Pfaff, Robert F. and Gilbert, Holly and Khodachenko, Maxim and Khomenko, Elena
and Shaikhislamov, Ildar F. and Soler, Roberto and V{\'a}zquez-Semadeni, Enrique and Zaqarashvili, Teimuraz},
title={Partially Ionized Plasmas in Astrophysics},
journal={Space Science Reviews},
year={2018},
month={Mar},
day={13},
volume={214},
number={2},
pages={58},
issn={1572-9672},
doi={10.1007/s11214-018-0485-6}
}

@article{jiang2024,
doi = {10.3847/1538-4357/ad9019},
year = {2024},
month = {dec},
publisher = {The American Astronomical Society},
volume = {977},
number = {2},
pages = {142},
author = {Jiang, Junnan and Liu, Yu and Huang, Wenlong and Li, Minchi and Jin, Rong and Yu, Pengcheng and Lei, Jiuhou},
title = {Plasma Waves Can Gain Energy from Neutrals in Partially Ionized Plasmas},
journal = {The Astrophysical Journal}
}

@book{kelley1989,
  author = {Kelley, M. C.},
  title = {The Earth's Ionosphere: Plasma Physics and Electrodynamics},
  publisher = {Elsevier},
  year = {1989},
  address = {Amsterdam}  
  }

@article{vernazza1981,
 author={Vernazza, J. E. and Avrett, E. H. and Loeser, R.},
 journal={Astrophysical Journal, Suppl. Ser.},
 volume={45},
 pages={635},
 year={1981},
 doi={10.1086/190731}
 }

@Article{lin2011,
author={Lin, Yong},
title={Filament Thread-like Structures and Their Small-amplitude Oscillations},
journal={Space Science Reviews},
year={2011},
month={Jul},
day={01},
volume={158},
number={2},
pages={237-266},
issn={1572-9672},
doi={10.1007/s11214-010-9672-9}
}

@article{diaz2012,
doi = {10.1088/0004-637X/754/1/41},
year = {2012},
month = {jul},
publisher = {The American Astronomical Society},
volume = {754},
number = {1},
pages = {41},
author = {Díaz, A. J. and Soler, R. and Ballester, J. L.},
title = {RAYLEIGH–TAYLOR INSTABILITY IN PARTIALLY IONIZED COMPRESSIBLE PLASMAS},
journal = {The Astrophysical Journal}
}

@article{parenti2024,
    author = {Parenti, S. and Luna, M. and Ballester, J. L.},
    title = {Future prospects for partially ionized solar plasmas: the prominence case},
    journal = {Philosophical Transactions of the Royal Society A: Mathematical, Physical and Engineering Sciences},
    volume = {382},
    number = {2272},
    pages = {20230225},
    year = {2024}, 
    month = {04},
    issn = {1364-503X},
    doi = {10.1098/rsta.2023.0225},
    url = {https://doi.org/10.1098/rsta.2023.0225}
}

@article{borah2007,
 title={Gravitational instability of partially ionized molecular clouds}, volume={73}, DOI={10.1017/S0022377806006295}, number={6}, journal={Journal of Plasma Physics}, author={BORAH, A.C. and SEN, A.K.}, year={2007},
pages={831–838}
 }

@ARTICLE{ballai2019,  
AUTHOR={Ballai, Istvan },     
TTLE={Linear Waves in Partially Ionized Plasmas in Ionization Non-equilibrium},       
JOURNAL={Frontiers in Astronomy and Space Sciences},      
VOLUME={6},
YEAR={2019},
DOI={10.3389/fspas.2019.00039},
ISSN={2296-987X}
}

@Article{kumar1990,
author={Kumar, Nagendra
and Srivastava, Krishna M.},
title={Gravitational instability of partially-ionized plasma carrying a uniform magnetic field with Hall effect},
journal={Astrophysics and Space Science},
year={1990},
month={Dec},
day={01},
volume={174},
number={2},
pages={211-216},
issn={1572-946X},
doi={10.1007/BF00642507}
}

@article{bhatia1995,
doi = {10.1088/0031-8949/51/6/012},
year = {1995},
month = {jun},
publisher = {},
volume = {51},
number = {6},
pages = {775},
author = {P K Bhatia and A B Rajib Hazarika},
title = {Gravitational instability of partially-ionized plasma in an oblique magnetic field},
journal = {Physica Scripta}
}

@article{rozina2023,
doi = {10.1088/1402-4896/acc218},
url = {https://doi.org/10.1088/1402-4896/acc218},
year = {2023},
month = {mar},
publisher = {IOP Publishing},
volume = {98},
number = {4},
pages = {045616},
author = {Rozina, Ch and Sania, B and Poedts, S and Ali, S and Maryam, N},
title = {Rayleigh-Taylor instability in an adiabatic-radiative rare plasma},
journal = {Physica Scripta}
}

@article{garai2020,
doi = {10.1088/1402-4896/abb697},
url = {https://doi.org/10.1088/1402-4896/abb697},
year = {2020},
month = {sep},
publisher = {IOP Publishing},
volume = {95},
number = {10},
pages = {105605},
author = {Garai, Sudip and Ghose-Choudhury, Anindya and Guha, Partha},
title = {Rayleigh Taylor like instability in presence of shear velocity in a strongly coupled quantum plasma},
journal = {Physica Scripta}
}

@article{rajaei2023,
doi = {10.1088/1402-4896/acba5f},
url = {https://doi.org/10.1088/1402-4896/acba5f},
year = {2023},
month = {mar},
publisher = {IOP Publishing},
volume = {98},
number = {4},
pages = {045604},
author = {Rajaei, Leila and Golpar-Raboky, Effat},
title = {The effect of collisions on the rayleigh-taylor instability in magnetized quantum plasma},
journal = {Physica Scripta}
}

@article{dey2025,
doi = {10.1088/1572-9494/ada3ca},
url = {https://doi.org/10.1088/1572-9494/ada3ca},
year = {2025},
month = {mar},
publisher = {IOP Publishing},
volume = {77},
number = {6},
pages = {065502},
author = {Dey, Rupak and Misra, A. P.},
title = {Rayleigh–Taylor instability in inhomogeneous relativistic classical and degenerate electron-ion magnetoplasmas},
journal = {Communications in Theoretical Physics}
}

@article{ruderman2018,
	author = {{Ruderman, M. S.} and {Ballai, I.} and {Khomenko, E.} and {Collados, M.}},
	title = {Rayleigh-Taylor instabilities with sheared magnetic fields  in partially ionised plasmas},
	DOI= {10.1051/0004-6361/201731534},
	journal = {Astronomy \& Astrophysics},
	year = {2018},
	volume = {609},
	pages = {A23}
}

@article{diaz2014,
	author = {{Díaz, A. J.} and {Khomenko, E.} and {Collados, M.}},
	title = {Rayleigh-Taylor instability in partially ionized compressible plasmas: One fluid approach},
	DOI= {10.1051/0004-6361/201322147},
	journal = {Astronomy \& Astrophysics},
	year = {2014},
	volume = {564},
	pages ={A97}
}

@article{lukin2024,
    author = {Lukin, Vyacheslav S. and Khomenko, Elena and Popescu Braileanu, Beatrice},
    title = {Mixing, heating and ion-neutral decoupling induced by Rayleigh–Taylor instability in prominence-corona transition regions},
    journal = {Philosophical Transactions of the Royal Society A: Mathematical, Physical and Engineering Sciences},
    volume = {382},
    number = {2272},
    pages = {20230417},
    year = {2024},
    month = {04},
    issn = {1364-503X},
    doi = {10.1098/rsta.2023.0417},
    url = {https://doi.org/10.1098/rsta.2023.0417}
   }

@article{hillier2023,
title = {Shocks and instabilities in the partially ionised solar atmosphere},
journal = {Advances in Space Research},
volume = {71},
number = {4},
pages = {1962-1983},
year = {2023},
note = {Recent progress in the physics of the Sun and heliosphere},
issn = {0273-1177},
doi = {https://doi.org/10.1016/j.asr.2022.08.079},
url = {https://www.sciencedirect.com/science/article/pii/S0273117722008158},
author = {Andrew Hillier and Ben Snow}
}

@article{soler2013,
  title={Magnetoacoustic waves in a partially ionized two-fluid plasma},
  author={Soler, Roberto and Carbonell, Marc and Ballester, Jose Luis},
  journal={The Astrophysical Journal Supplement Series},
  volume={209},
  number={1},
  pages={16},
  year={2013},
  publisher={IOP Publishing}
}

@article{vranjes2015,
    author = {Vranjes, J. and Kono, M. and Luna, M.},
    title = {Charge exchange in fluid description of partially ionized plasmas},
    journal = {Monthly Notices of the Royal Astronomical Society},
    volume = {455},
    number = {4},
    pages = {3901-3909},
    year = {2015},
    month = {12},
    issn = {0035-8711},
    doi = {10.1093/mnras/stv2595},
    url = {https://doi.org/10.1093/mnras/stv2595}
   }

@article{song2017,
doi = {10.3847/1538-4357/aa85e1},
url = {https://doi.org/10.3847/1538-4357/aa85e1},
year = {2017},
month = {sep},
publisher = {The American Astronomical Society},
volume = {846},
number = {2},
pages = {92},
author = {Song, P.},
title = {A Model of the Solar Chromosphere: Structure and Internal Circulation},
journal = {The Astrophysical Journal}
}

@article{braileanu2019,
	author = {{Popescu Braileanu, B.} and {Lukin, V. S.} and {Khomenko, E.} and {de Vicente, Á.}},
	title = {Two-fluid simulations of waves in the solar chromosphere - II. Propagation and damping of fast magneto-acoustic waves and shocks},
	DOI= {10.1051/0004-6361/201935844},
	journal = {A \& A},
	year = {2019},
	volume = {630},
	pages = {A79}
}

@article{vranjes2013,
	author = {{Vranjes, J.} and {Krstic, P. S.}},
	title = {Collisions, magnetization, and transport coefficients in the lower solar atmosphere},
	DOI= {10.1051/0004-6361/201220738},
	url= {https://doi.org/10.1051/0004-6361/201220738},
	journal = {A \& A},
	year = {2013},
	volume = {554},
	pages = {A22} 	 
}
\nopagebreak
\end{document}